\documentclass[12pt]{article}
\usepackage{amsfonts}
\usepackage[latin5]{inputenc}
\usepackage{graphicx}
\usepackage{amsmath}
\usepackage{amssymb}
\usepackage{xypic}

\begin{document}
\begin{titlepage}
\title{Degenerate Spin Structures and the L\'{e}vy-Leblond Equation}
\author{ T. Dereli\thanks{E. mail:
tdereli@ku.edu.tr}\\
\small{Department of Physics, Ko\c{c} University}\\
\small{34450 Sar{\i}yer-\.{I}stanbul, Turkey}
\\
\\
\c{S}. Ko\c{c}ak\thanks{E. mail: skocak@anadolu.edu.tr} \, , \, M.
Limoncu\thanks{E. mail: mlimoncu@anadolu.edu.tr}
\\ {\small Department of Mathematics,
 Anadolu University}\\ \small{26470 Eski\c{s}ehir, Turkey}}
\date{ }
\maketitle

\begin{abstract}
\noindent Newton-Cartan manifolds and the Galilei group are
defined by the use of co-rank one degenerate metric tensor.
Newton-Cartan connection is lifted to the degenerate spinor bundle
over a Newton-Cartan 4-manifold by the aid of degenerate spin
group. Levy-Leblond equation is constructed with the lifted
connection.
\end{abstract}

\bigskip

\noindent {\small Paper presented at the {\bf 5th Workshop on
Quantization, Dualities and Integrable Systems}, Pamukkale
University, 23-28 January 2006, Denizli, Turkey.}
\end{titlepage}

\section{Motivation}

Degenerate spin structures in general is a subject that received
little attention up to now.  This  provides by itself sufficient
reason to study degenerate Clifford algebras and related
structures in differential geometry.

From  physics point of view too this is an interesting but a
neglected subject. It is well-known that Newton's non-relativistic
theory of gravitation may be given a locally Galilei covariant
formulation over a 4-dimensional space-time equipped with
independent space and time metrics that are both degenerate. A
Newton-Cartan manifold is a space-time that admits a linear
connection compatible with both space and time metrics.

In what follows, the degenerate Clifford algebra
$\mathcal{C}\ell_{1,0,3}$ is defined over a 4-dimensional
Newton-Cartan manifold. The corresponding degenerate spin group
$SPIN(1,0,3)$ is defined. The Newton-Cartan connection is lifted
to the degenerate spinor bundle. This allows one to write down the
L\`{e}vy-Leblond equation satisfied by a non-relativistic spin-1/2
electron directly, thus coupling it to gravity.

\section{Galilei Group and Degenerate Spin Group}

Let $\langle,\rangle$ be a symmetric bilinear form on
$\mathbb{R}^{n}$ and let us consider the subspace $W$ of
$\mathbb{R}^{n}$
\begin{eqnarray*}
W=\{w\in\mathbb{R}^{n}\mid\langle w,x\rangle=0 \text{ for all }
x\in\mathbb{R}^{n}\}.
\end{eqnarray*}
$W$ is called the radical and the dimension of $W$ is called the
co-rank of $\langle,\rangle$. If co-rank is zero/non-zero, then
$\langle,\rangle$ is called non-degenerate/degenerate. If
$W'\subset\mathbb{R}^{n}$ is any complementary subspace to $W$,
then the restriction of $\langle,\rangle$ to $W'$ is
non-degenerate \cite{DB}. If the co-rank of $\langle,\rangle$ is
$r$ and the the restriction of $\langle,\rangle$ to $W'$ has
signature $(p,q)$ (in the sense that $W'$ can be decomposed as
$W''\oplus W'''$, where the dimension of $W''$ respectively $W'''$
are $p$ resp. $q$ and the restriction of $\langle,\rangle$ to
$W''$ resp. $W'''$ is negative resp. positive definite), then the
pair $(\mathbb{R}^{n},\langle,\rangle)$ is said to be of type
$\mathbb{R}^{r,p,q}$. In this terminology, $\mathbb{R}^{0,1,3}$ is
called relativistic space-time and $\mathbb{R}^{1,0,3}$ is called
non-relativistic space-time.

We will consider throughout the non-relativistic
$\mathbb{R}^{1,0,3}$. Let the radical be spanned by the vector $f$
and let $\{e_{1},e_{2},e_{3}\}$ be an orthonormal basis for a
complementary $W'$. Then, obviously, the set
$\{f,e_{1},e_{2},e_{3}\}$ becomes a basis for
$\mathbb{R}^{1,0,3}$.

In a non-relativistic space-time, the set of linear-automorphisms
$\varphi$ from $\mathbb{R}^{1,0,3}$ to $\mathbb{R}^{1,0,3}$ is
called the Galilei group if
$\langle\varphi(v),\varphi(w)\rangle=\langle v,w\rangle$ for all
$v,w\in\mathbb{R}^{1,0,3}$, $det(\varphi)=1$ and
$\varphi\!\!\mid_{W}=id$ . Thus the Galilei group can be written
as
\begin{eqnarray*}
SO(1,0,3)=\{\varphi\in Aut(\mathbb{R}^{1,0,3})\mid
\langle\varphi(v),\varphi(w)\rangle=\langle
v,w\rangle,\hspace{0.1cm}det(\varphi)=1,\hspace{0.1cm}\varphi\!\!\mid_{W}=id\}.
\end{eqnarray*}
Using the matrix of $\varphi$ denoted by $\Phi$ with respect to
the basis $\{f,e_{1},e_{2},e_{3}\}$, we can write
\begin{eqnarray*}
SO(1,0,3)=\{\Phi\in Mat(4\times4)\mid
\Phi^{t}G\Phi=G,\hspace{0.1cm}det(\Phi)=1,\hspace{0.1cm}\Phi\!\mid_{sp\{f\}}=I\}
\end{eqnarray*}
where $W=sp\{f\}=\{\lambda f\mid\lambda\in\mathbb{R}\}$ and
\begin{eqnarray*}
G=\left(\begin{array}{cc}0 & 0\\0 &
\,\,\,\,\,\,I_{3\times3}\end{array}\right).
\end{eqnarray*}
From the above equations we obtain that $SO(1,0,3)$ can be
parameterized as
\begin{eqnarray*}
SO(1,0,3)=\{\left(\begin{array}{cc}1 & A\\0 &
R\end{array}\right)\mid R\in SO(3)\hspace{0.3cm} A\in
Mat(1\times3)\}.
\end{eqnarray*}

Moreover $SO(1,0,3)$ is also isomorphic to the semi-direct product
of $SO(3)$ and $Mat(3\times1)$
\begin{eqnarray*}
SO(1,0,3)\simeq SO(3)\ltimes_{id}Mat(3\times1).
\end{eqnarray*}

Let $\mathcal{C}\ell_{1,0,3}$ be the degenerate Clifford algebra
of $\mathbb{R}^{1,0,3}$. Degenerate spin group is defined to be a
special subset of $\mathcal{C}\ell_{1,0,3}$
\begin{eqnarray*}
SPIN(1,0,3)=\{s(1+vf)\mid s\in SPIN(3),v\in
W'=sp\{e_{1},e_{2},e_{3}\}\}
\end{eqnarray*}
where $SPIN(3)$ is taken as the spin group lying in the Clifford
algebra on $W'$,
$\mathcal{C}\ell(W')\subset\mathcal{C}\ell_{1,0,3}$. We can regard
$SPIN(3)$ as a subspace of $SPIN(1,0,3)$ taking $v=\bold{0}$. (For
the general $SPIN(r,p,q)$ see \cite{C})

$SPIN(1,0,3)$ is indeed a group under the Clifford multiplication.
Let us first see that the Clifford multiplication on $SPIN(1,0,3)$
is a binary-operation:
\begin{eqnarray*}
s(1+vf)s'(1+v'f)=(s+svf)(s'+s'v'f)\\
=ss'+ss'v'f+svfs'+svfs'v'f\hspace{-1.870cm}\\
=ss'+ss'v'f+svs'f-svf^{2}s'v'\hspace{-1.79cm}\\
=ss'+ss'v'f+svs'f\hspace{0.15cm}\\
=ss'+ss'v'f+ss's'^{-1}vs'f\hspace{-0.85cm}
\end{eqnarray*}
by $f^{2}=0$. Using the 2:1 group homomorphism
$\rho:SPIN(3)\rightarrow SO(3)$ which is defined by
$s\mapsto\rho(s)(v)=svs^{-1}$, we can write
\begin{eqnarray*}
s(1+vf)s'(1+v'f)=ss'+ss'v'f+ss'\rho(s'^{-1})(v)f\\
=ss'(1+(v'+\rho(s'^{-1})(v))f).\hspace{0.26cm}
\end{eqnarray*}
\noindent Since the Clifford algebra is an associative algebra,
the Clifford multiplication is also associative on $SPIN(1,0,3)$.

\noindent Obviously, $1\in\mathcal{C}\ell_{1,0,3}$ belongs to
$SPIN(1,0,3)$.

\noindent Inverse of any element of $SPIN(1,0,3)$ is given by
\begin{eqnarray*}
(s(1+vf))^{-1}=s^{-1}(1-\rho(s)(v)f).
\end{eqnarray*}

Degenerate spin group can also be interpreted as a semi-direct
product \cite{K},\cite{M}
\begin{eqnarray*}
SPIN(1,0,3)\simeq SPIN(3)\ltimes_{\rho}\mathbb{R}^{3}
\end{eqnarray*}
where $\rho:SPIN(3)\rightarrow SO(3)$ is the 2:1 group
homomorphism.

\vskip 5mm

There is a second 2:1 group homomorphism given by
\begin{eqnarray*}
\begin{array}{cccc}
\rho': & SPIN(1,0,3) & \longrightarrow & SO(1,0,3)\\
\, & \frak{s} & \longmapsto &
\rho'(\frak{s})(x)=\frak{s}x\frak{s}^{-1}
\end{array}
\end{eqnarray*}
or equally,
\begin{eqnarray*}
\begin{array}{cccc}
\rho': & SPIN(1,0,3) & \longrightarrow & SO(1,0,3)\\
\, & \frak{s}=s(1+vf) & \longmapsto & \left(\begin{array}{cc}1 &
2v\\0&\rho(s)\end{array}\right)
\end{array}
\end{eqnarray*}
where the vector $v=v_{1}e_{1}+v_{2}e_{2}+v_{3}e_{3}$ is taken to
be a row vector. \vskip 5mm

Lie algebra of $SO(1,0,3)$ is
\begin{eqnarray*}
so(1,0,3)=\{\phi\in Mat(4\times4)\mid (G\phi)^{t}+G\phi=0\}
\end{eqnarray*}
that is parameterized as
\begin{eqnarray*}
so(1,0,3)=\{\left(\begin{array}{cc}0 & a\\0 &
r\end{array}\right)\mid r\in so(3)\hspace{0.3cm} a\in
Mat(1\times3)\}.
\end{eqnarray*}
\vskip 5mm

After this we will consider the following basis for $so(1,0,3)$
\begin{eqnarray*}
E_{01}=\left(\begin{array}{cccc}0&1&0&0\\0&0&0&0\\0&0&0&0\\0&0&0&0\end{array}\right)
E_{02}=\left(\begin{array}{cccc}0&0&1&0\\0&0&0&0\\0&0&0&0\\0&0&0&0\end{array}\right)
E_{03}=\left(\begin{array}{cccc}0&0&0&1\\0&0&0&0\\0&0&0&0\\0&0&0&0\end{array}\right)
\end{eqnarray*}
\begin{eqnarray*}
E_{12}=\left(\begin{array}{cccc}0&0&0&0\\0&0&1&0\\0&-1&0&0\\0&0&0&0\end{array}\right)
E_{13}=\left(\begin{array}{cccc}0&0&0&0\\0&0&0&1\\0&0&0&0\\0&-1&0&0\end{array}\right)
E_{23}=\left(\begin{array}{cccc}0&0&0&0\\0&0&0&0\\0&0&0&1\\0&0&-1&0\end{array}\right)
\end{eqnarray*}
\vskip 5mm

Lie algebra of $SPIN(1,0,3)$ is given by the span of the following
elements of $\mathcal{C}\ell_{1,0,3}$
\begin{eqnarray*}
spin(1,0,3)=\{a\mid a\in
sp\{fe_{1},fe_{2},fe_{3},e_{1}e_{2},e_{1}e_{3},e_{2}e_{3}\}\}\subset
\mathcal{C}\ell_{1,0,3}.
\end{eqnarray*}

Differential map
$(d\rho')_{1_{SPIN(1,0,3)}}:spin(1,0,3)\longrightarrow so(1,0,3)$
is an isomorphism between Lie algebras $spin(1,0,3)$ and
$so(1,0,3)$. Their basis elements correspond as follows:
\begin{eqnarray*}
(d\rho')_{1_{SPIN(1,0,3)}}(e_{i}e_{j})=2E_{ij}\hspace{-5.1cm}\\
&(d\rho')_{1_{SPIN(1,0,3)}}(fe_{i})=-2E_{0i}
\end{eqnarray*}

\section{Newton-Cartan Manifold, Degenerate Spin Manifold and Degenerate Spinor Bundle}
Let $g$ be a co-rank one degenerate symmetric metric tensor field
on smooth 4-manifold $M$, and let $\tau$ be a smooth 1-form such
that $\tau(V)=1$ where the non-zero smooth vector field $V$
satisfies $g(X,V)=0$ for all vector fields $X$. Then we obtain the
following tensor fields (\cite{DKL}): \vskip 3mm \noindent
$\bar{g}=\tau\otimes\tau+g$ is a non-degenerate symmetric
(0,2)-tensor field. In fact, $\bar{g}$ is obviously symmetric. For
non-degeneracy, let us consider kernel of $\mathbb{R}$-linear map
$\tau_{p}:T_{p}M\rightarrow \mathbb{R}$ for all $p\in M$. As
$ker(\tau_{p})=\{U_{p}\in T_{p}M|\tau_{p}(U_{p})=0\}$ is a
subspace of $T_{p}M$ and
$dim(T_{p}M)=dim(ker(\tau_{p}))+dim(\mathbb{R})$, the dimension of
$ker(\tau_{p})$ is found to be $3$. From $\tau(V)=1$ we see that
$V_{p}$ is not element of $ker(\tau_{p})$ because
$\tau_{p}(V_{p})=1$. Then choosing a basis
$\{(U_{1})_{p},(U_{2})_{p},(U_{3})_{p}\}$ of $ker(\tau_{p})$, the
set $\{V_{p},(U_{1})_{p},(U_{2})_{p},(U_{3})_{p}\}$ becomes a
basis of $T_{p}M$. On the other hand $g_{p}$ restricted to
$ker(\tau_{p})$ is non-degenerate because of $rank(g)=3$.
Otherwise from the basis
$\{V_{p},(U_{1})_{p},(U_{2})_{p},(U_{3})_{p}\}$ we would find
$rank(g)<3$. The basis $\{(U_{1})_{p},(U_{2})_{p},(U_{3})_{p}\}$
of $ker(\tau_{p})$ can be chosen orthonormal. Finally if we
construct the matrix of $\bar{g}$ with respect to the basis
$\{V_{p},(U_{1})_{p},(U_{2})_{p},(U_{3})_{p}\}$ we see that
\begin{eqnarray*}
\bar{g}_{p}(V_{p},V_{p})=\tau_{p}(V_{p})\tau_{p}(V_{p})+g_{p}(V_{p},V_{p})=1\hspace{3.25cm}\\
\bar{g}_{p}(V_{p},(U_{i})_{p})=\tau_{p}(V_{p})\tau_{p}((U_{i})_{p})+g_{p}(V_{p},(U_{i})_{p})=0\hspace{1.8cm}\\
\bar{g}_{p}((U_{i})_{p},(U_{j})_{p})=\tau_{p}((U_{i})_{p})\tau_{p}((U_{j})_{p})
+g_{p}((U_{i})_{p},(U_{j})_{p})=\delta_{ij}
\end{eqnarray*}
the matrix is diagonal and $det(\bar{g}_{p})=1\neq0$ for all $p\in
M$ where $i,j=1,2,3$. Therefore $\bar{g}_{p}$ is non-degenerate.
Note that non-degeneracy (and degeneracy) is independent of the
chosen of basis.
\newline\noindent
Secondly there is another non-degenerate symmetric (2,0)-tensor
field $\bar{h}$ given by
$\bar{h}(\alpha,\beta)=\bar{g}(\alpha^{*},\beta^{*})$ for all
one-forms $\alpha,\beta$, where the star $*$ denotes the
metric-dual of one-forms, and we have, locally, the relation
$\bar{h}^{\mu\lambda}\bar{g}_{\lambda\nu}=\delta^{\mu}_{\nu}$.\newline
\textit{Note}: For each one -form $\alpha$ there exits a unique
vector field $\alpha^{*}$, metric-dual of one-form $\alpha$, such
that $\alpha(X)=\bar{g}(\alpha^{*},X)$ for all vector fields $X$.
Therefore we have an isomorphism from the set of one-forms to the
set of vector fields $\alpha\longmapsto\alpha^{*}$. As known, the
set of one-forms and vector fields are modules over the
commutative ring $C^{\infty}(M,\mathbb{R})$.\newline $\bar{h}$ is
obviously symmetric and non-degenerate because of the above note
and non-degeneracy of $\bar{g}$. The local relation is obtained as
follows: In any coordinate system, if we take $\alpha=dx^{\eta}$,
then we obtain
$(dx^{\mu})^{*}=(\bar{g}^{\,-1})^{\mu\kappa}\partial_{\kappa}$ by
the aid of $\alpha(X)=\bar{g}(\alpha^{*},X)$ (where
$(\bar{g}^{\,-1})^{\mu\nu}$ is inverse matrix of
$\bar{g}_{\mu\nu}$). Thus we have
\begin{eqnarray*}
\bar{h}(dx^{\mu},dx^{\nu})=\bar{g}((dx^{\mu})^{*},(dx^{\nu})^{*})\\
\bar{h}^{\mu\nu}=(\bar{g}^{\,-1})^{\mu\kappa}(\bar{g}^{\,-1})^{\nu\sigma}\bar{g}_{\kappa\sigma}
=(\bar{g}^{\,-1})^{\nu\mu}\hspace{-0.8cm}
\end{eqnarray*}
and from the symmetry of $\bar{g}^{\,-1}$
\begin{eqnarray*}
\bar{h}^{\mu\nu}=(\bar{g}^{\,-1})^{\mu\nu}.
\end{eqnarray*}
Therefore the expression
$\bar{h}^{\mu\lambda}\bar{g}_{\lambda\nu}$ equals to
$\delta^{\mu}_{\nu}$.
\newline
Finally, we note that $h=-V\otimes V+\bar{h}$ is a co-rank one
degenerate symmetric (2,0)-tensor field such that
$h^{\mu\lambda}g_{\lambda\nu}=\delta^{\mu}_{\nu}-V^{\mu}\tau_{\nu}$.

$M$ is called a Newton-Cartan manifold if $M$ can be furnished by
tensor fields $g$ and $\tau$ satisfying the above properties. A
linear connection on $M$ compatible with $g$ and $\tau$ will be
called a Newton-Cartan connection \cite{DKL}.

Let $M$ be a Newton-Cartan manifold, and let us consider the
principal bundle $P_{SO(1,0,3)}$ on $M$. Then $M$ will be called a
degenerate spin manifold if there exists a principal $SPIN(1,0,3)$
bundle $P_{SPIN(1,0,3)}$ on $M$ satisfying the following property:
Transition maps $\varphi_{\alpha\beta}$ and
$\widetilde{\varphi}_{\alpha\beta}$ of the principal bundles
$P_{SO(1,0,3)}$ and $P_{SPIN(1,0,3)}$ can be chosen on a joint
covering $\{U_{\alpha}\}$, such that the following diagram
commutes:

\begin{displaymath}
\xymatrix{\, & SPIN(1,0,3) \ar[d]_{\rho'}^{2:1}\\
U_{\alpha}\cap U_{\beta}
\ar[ur]^{\widetilde{\varphi}_{\alpha\beta}}
\ar[r]_{\varphi_{\alpha\beta}} & SO(1,0,3)}
\end{displaymath}

An algebra homomorphism
$\theta:\mathcal{C}\ell_{1,0,3}\longrightarrow
End(\mathbb{C}^{4})$, i.e. a matrix representation of the
degenerate Clifford algebra $\mathcal{C}\ell_{1,0,3}$, gives a
group homomorphism $\theta:SPIN(1,0,3)\longrightarrow
Aut(\mathbb{C}^{4})$. By the aid of this group homomorphism we can
construct the associated vector bundle
$P_{SPIN(1,0,3)}\times_{\theta} \mathbb{C}^{4}$ which is called a
degenerate spinor bundle.

\section{Lifting the Connection}
Let $\nabla$ be a linear connection on $M$ such that $\nabla_{Z}
g=0$ for all vector fields $Z$, i.e., $g$-compatible linear
connection. We define the connection 1-form on $P_{SO(1,0,3)}$ by
\begin{eqnarray*}
\mathcal{A}_{\alpha}(W)=e^{a}(\nabla_{W}X_{b})=\sum_{c=0}^{3}W^{c}\Gamma_{cb}^{a}
\end{eqnarray*}
or
\begin{eqnarray*}
\mathcal{A}_{\alpha}=\sum_{c=0}^{3}\Gamma_{cb}^{a}e^{c}=\omega_{\,\,\,b}^{a}
\end{eqnarray*}
on $U_{\alpha}\subset M$. Here $\{e^{a}\}$, $\{X_{a}\}$ are local
coframe and frame field, $W\in \Gamma(TU_{\alpha})$ and
$\mathcal{A}_{\alpha}(W)(x)\in so(1,0,3)$ for all $x\in
U_{\alpha}$. Using the basis of $so(1,0,3)$, we can write
\begin{eqnarray*}
\mathcal{A}_{\alpha}(W)=\sum_{i=1}^{3}\omega_{\,\,\,i}^{0}(W)\,E_{0i}+\sum_{i<j=1}^{3}\omega_{\,j}^{i}(W)\,E_{ij}.
\end{eqnarray*}

Since $M$ is a spin manifold, $\mathcal{A}_{\alpha}$ can be lifted
to $P_{SPIN(1,0,3)}$ as
\begin{eqnarray*}
\widetilde{\mathcal{A}}_{\alpha}(W)=(d\rho')^{-1}_{1_{SPIN(1,0,3)}}(\mathcal{A}_{\alpha}(W)).
\end{eqnarray*}
From this expression we obtain
\begin{eqnarray*}
\widetilde{\mathcal{A}}_{\alpha}(W)&=&-\frac{1}{2}\sum_{i=1}^{3}\omega_{\,\,\,i}^{0}(W)fe_{i}
+\frac{1}{2}\sum_{i<j=1}^{3}\omega_{\,j}^{i}(W)e_{i}e_{j}
\end{eqnarray*}

Secondly $\widetilde{\mathcal{A}}_{\alpha}$ is lifted to the
degenerate spinor bundle $P_{SPIN(1,0,3)}\times_{\theta}
\mathbb{C}^{4}$ as
\begin{eqnarray*}
\overline{\mathcal{A}}_{\alpha}(W)=(d\theta)_{1_{SPIN(1,0,3)}}(\widetilde{\mathcal{A}}_{\alpha}(W))
\end{eqnarray*}
by the differential map
$(d\theta)_{1_{SPIN(1,0,3)}}:spin(1,0,3)\rightarrow
gl(4,\mathbb{C})$ which is given by
\begin{eqnarray*}
(d\theta)_{1_{SPIN(1,0,3)}}(e_{i}e_{j})=\theta(e_{i})\theta(e_{j})\\
(d\theta)_{1_{SPIN(1,0,3)}}(fe_{i})=\theta(f)\theta(e_{i}).
\end{eqnarray*}
Thus one gets
\begin{eqnarray*}
\overline{\mathcal{A}}_{\alpha}(W)=-\frac{1}{2}\sum_{i=1}^{3}\omega_{\,\,\,i}^{0}(W)\theta(f)\theta(e_{i})
+\frac{1}{2}\sum_{i<j=1}^{3}\omega_{\,j}^{i}(W)\theta(e_{i})\theta(e_{j}).
\end{eqnarray*}

Using $\overline{\mathcal{A}}_{\alpha}$, the following connection
is defined on the degenerate spinor bundle:
\begin{eqnarray*}
(\nabla\psi_{\alpha})(W)=\nabla_{W}\psi_{\alpha}
=(d\psi_{\alpha})(W)+(\overline{\mathcal{A}}_{\alpha}(W))(\psi_{\alpha})
\end{eqnarray*}
and
\begin{equation*}
\begin{array}{c}
(\nabla\psi_{\alpha})(W)=\nabla_{W}\psi_{\alpha}
=(d\psi_{\alpha})(W)
-\displaystyle\frac{1}{2}\displaystyle\sum_{i=1}^{3}\omega_{\,\,\,i}^{0}(W)\theta(f)\theta(e_{i})(\psi_{\alpha})\\
\hspace{6.2cm}+\displaystyle\frac{1}{2}\displaystyle\sum_{i<j=1}^{3}
\omega_{\,j}^{i}(W)\theta(e_{i})\theta(e_{j})(\psi_{\alpha}).
\end{array}
\end{equation*}

\section{L\`{e}vy-Leblond Equation}
By the use of the equivalent bundles
$P_{SPIN(1,0,3)}\times_{\sigma}\mathcal{C}\ell_{1,0,3}\cong\mathcal{C}\ell(TM)$,
where $\sigma:SPIN(1,0,3)\longrightarrow
Aut(\mathcal{C}\ell_{1,0,3})$ is the group homomorphism given by
$\frak{s}\longmapsto \sigma(\frak{s})(c):=\frak{s}c\frak{s}^{-1}$,
the algebra homomorphism
$\theta:\mathcal{C}\ell_{1,0,3}\longrightarrow
End(\mathbb{C}^{4})$ can be extended as follows:
\begin{eqnarray*}
\begin{array}{cccc}
\theta: &
P_{SPIN(1,0,3)}\times_{\sigma}\mathcal{C}\ell_{1,0,3}\cong\mathcal{C}\ell(TM)
&\longrightarrow &End(P_{SPIN(1,0,3)}\times_{\theta}
\mathbb{C}^{4})\\
\,& [u,c]&\longmapsto & \theta([u,c])([u,v]):=[u,\theta(c)(v)]
\end{array}
\end{eqnarray*}
Using this extension we define the operator
\begin{eqnarray*}
D:\Gamma(P_{SPIN(1,0,3)}\times_{\theta}
\mathbb{C}^{4})\longrightarrow
\Gamma(P_{SPIN(1,0,3)}\times_{\theta} \mathbb{C}^{4})
\end{eqnarray*}
by
\begin{eqnarray*}
D\psi=\sum_{a,b=0}^{3}\bar{h}(e^{a},e^{b})\theta(X_{a})((\nabla\psi)(X_{b}))
=\sum_{a,b=0}^{3}\bar{h}(e^{a},e^{b})\theta(X_{a})(\nabla_{X_{b}}\psi),
\end{eqnarray*}
where $\bar{h}$ is the non-degenerate contravariant tensor field
on the Newton-Cartan manifold. This operator can be written as
\begin{eqnarray*}
D\psi=\sum_{a=0}^{3}\gamma^{a}(\nabla\psi)(X_{a})=\sum_{a=0}^{3}\gamma^{a}\nabla_{X_{a}}\psi
\end{eqnarray*}
with the notation
\begin{eqnarray*}
\bar{h}(e^{a},e^{b})=\bar{h}^{ab}\qquad\theta(X_{a})=\gamma_{a}\qquad\bar{h}^{ab}\gamma_{b}=\gamma^{a}.
\end{eqnarray*}
Using the expression of $\nabla_{X_{a}}\psi$, we find
\begin{eqnarray*}
\begin{array}{c}
D\psi=\displaystyle\sum_{a=0}^{3}\gamma^{a}((d\psi_{\alpha})(X_{a})
-\displaystyle\frac{1}{2}\displaystyle\sum_{i=1}^{3}\omega_{\,\,\,i}^{0}(X_{a})\theta(f)\theta(e_{i})(\psi_{\alpha})\\
\hspace{4.9cm}+\displaystyle\frac{1}{2}\displaystyle\sum_{i<j=1}^{3}
\omega_{\,j}^{i}(X_{a})\theta(e_{i})\theta(e_{j})(\psi_{\alpha}))
\end{array}
\end{eqnarray*}
or
\begin{eqnarray*}
\begin{array}{c}
D\psi=\displaystyle\sum_{a=0}^{3}\gamma^{a}((d\psi_{\alpha})(X_{a})
-\displaystyle\frac{1}{2}\displaystyle\sum_{i=1}^{3}\Gamma^{0}_{ai}\theta(f)\theta(e_{i})(\psi_{\alpha})\\
\hspace{4.9cm}+\displaystyle\frac{1}{2}\displaystyle\sum_{i<j=1}^{3}
\Gamma^{i}_{aj}\theta(e_{i})\theta(e_{j})(\psi_{\alpha})).
\end{array}
\end{eqnarray*}
Under the following representation of $\mathcal{C}\ell_{1,0,3}$,
\begin{eqnarray*}
\theta(f)=\gamma_{0}=\left(\begin{array}{cc}
0&I\\
0&0\end{array} \right)\qquad
\theta(e_{i})=\gamma_{i}=\left(\begin{array}{cc}
\sigma_{i}&0\\
0&-\sigma_{i}\end{array} \right)
\end{eqnarray*}
where $\sigma_{i}$ are the Pauli spin matrices, we obtain the
equation
\begin{eqnarray*}
D\psi+2mi\gamma_{0}^{t}\psi=0,
\end{eqnarray*}
which is called the L\`{e}vy-Leblond equation \cite{LL},\cite{K}.
In the flat case ($X_{a}=\partial_{a}$, $e^{a}=dx^{a}$,
$\Gamma^{c}_{ab}=0$) it becomes the original L\`{e}vy-Leblond
equation
\begin{eqnarray*}
\gamma^{a}\partial_{a}\psi+2mi\gamma_{0}^{t}\psi=0
\end{eqnarray*}
and from this one gets by iteration the Schrödinger equation
\begin{eqnarray*}
-\frac{1}{2m} \triangle = i\partial_{t}\psi .
\end{eqnarray*}

\end{document}